# A Practical Dirty Paper Coding Applicable for Broadcast Channel

Srikanth B. Pai

Coding and Modulation Lab, Dept of ECE,
Indian Institute of Science,
Bangalore 560012, India
Email:spai@ece.iisc.ernet.in

B. Sundar Rajan

Coding and Modulation Lab, Dept of ECE,
Indian Institute of Science,
Bangalore 560012, India
Email:bsrajan@ece.iisc.ernet.in

*Abstract*—In this paper, we present a practical dirty paper coding scheme using trellis coded modulation for the dirty paper channel $Y = X+S+W$, $\mathbb{E}\{X^2\} \leq P$, where $W$ is white Gaussian noise with power $\sigma_w^2$, $P$ is the average transmit power and $S$ is the Gaussian interference with power $\sigma_s^2$ that is non-causally known at the transmitter. We ensure that the dirt in our scheme remains distinguishable to the receiver and thus, our designed scheme is applicable to broadcast channel. Following Costa's idea, we recognize the criteria that the transmit signal must be as orthogonal to the dirt as possible. Finite constellation codes are constructed using trellis coded modulation and by using a Viterbi algorithm at the encoder so that the code satisfies the design criteria and simulation results are presented with codes constructed via trellis coded modulation using QAM signal sets to illustrate our results.

## I. INTRODUCTION

The dirty paper channel model is receiving increased attention due to its utility in communication over Gaussian broadcast channels and design of good methods for digital watermarking and information embedding. In this communication channel model, characterised by

$$Y = X + S + W, \; \mathbb{E}(X^2) \leq P \qquad (1)$$

a Gaussian interference $S$, whose realizations are non-causally known to the transmitter (but not to the receiver), is added to the transmitted signal $X$ along with white Gaussian noise $W$ as shown by Fig. 1. It is assumed that the noise is independent of the interference and transmit signal. Costa, in 1983, proved that the capacity of this channel is the same as the capacity of the dirty paper channel without the interference [1]. This result is, perhaps, surprising because it suggests that there is no loss in power, as for as the capacity is concerned, when the transmitter *adapts* to the interfering signal.

It was shown that the two user Gaussian broadcast capacity for the point to point case can be achieved by the use of dirty paper coding in [6]. The two user broadcast problem is that a transmitter wants to convey a separate signal $X_{m_1}$ and $X_{m_2}$ reliably to each of it's two receivers simultaneously. We will term the user with less noise variance (say user 1) the "strong user" and the remaining user the "weak user" and choose to add the codewords of these two users (i.e. the transmit signal $X = X_{m_1} + X_{m_2}$) to explain their idea. The transmitter chooses a Gaussian codebook with a fraction of allowed transmit power for the weaker user. The codeword of the weaker user $\mathbf{X}(m_2)$ is non-causally known to the transmitter. Since it was decided that we will add the codewords of the weaker user and the broadcasting transmitter knows the weaker user's codewords, the channel for the stronger user will look like a dirty paper channel. That is, the channel looks like $\mathbf{Y_1} = \mathbf{X}(m_1) + \mathbf{X}(m_2) + \mathbf{W_1}$ and while coding for the strong user, the additive interference $\mathbf{X}(m_2)$ is known to the transmitter. The authors in [6] show that there is no loss in rate when the stronger user performs dirty paper coding, by achieving the well known capacity of the 2 user Gaussian broadcast region. In the dirty paper coding problem, our intentions are only to transmit the message across the channel without worrying about the status of the dirt at the receiver. It is important to observe, at this point, that the *dirt in the case of the broadcast channel is the weaker user's information and protecting it's integrity is crucial to communicating over the broadcast channel.* Thus a good solution to the dirty paper coding problem that reduces the distinguishability of the dirt may not be a good solution to the broadcast coding problem. A similar argument holds for digital watermarking also [8].

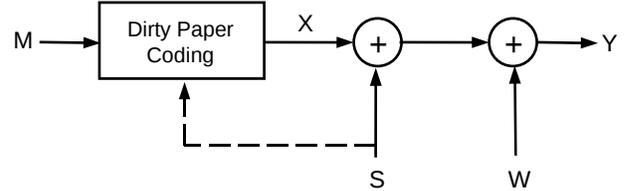

Fig. 1. A Gaussian Dirty Paper Channel

A large amount of work has been done in designing structured codebooks which achieve Costa's promised rate asymptotically (in block length). Zamir *et al* quantize the interference into bins and then use a capacity achieving additive white Gaussian noise (AWGN) code for each bin [10]. They use lattices as the framework for this idea and postulate the *existence* of a good lattice. This approach is pursued from a practical view point in [11]. For the case of finite alphabets, Gariby *et al* have explored bounds on the capacity of the dirty paper channel, assuming a Pulse Amplitude Modulated (PAM) constellation [14]. All these approaches attempt to find the maximum, reliable rate of transmission with a constrained probability of error. Recently, Skoglund and Larsson have investigated optimal modulation schemes for the dirty paper channel [13], from the view point of error performance for a fixed rate. They optimize the modulator and the demodulator jointly for the case of a one-dimensional modulation scheme. The design involves solving an optimization problem for every value of the interference. Since it is not feasible for a modulator to solve it, they quantize the interference and construct a lookup table for different values of interfering signals.

The most popular finite constellation coding approach to dirty paper coding uses Tomlinson-Harashima precoding (THP) introduced in [2],[3]. Originally, THP was a single dimensional pre-equalization technique designed to combat intersymbol interference (ISI). The basic idea of THP is to subtract interference, quantize the subtracted signal so that the power is constrained within the required value. The received signal is then decoded with respect to the quantized framework. The main motivation to use THP in dirty paper coding is to design a coding scheme for the dirty paper channel that performs independent of the dirt power in an attempt to reinforce Costa's capacity result. In [9], the authors pointed out the connections between THP and the dirty paper result. They also introduce the idea of partial interference presubtraction (PIP) and show that it outperforms completely subtracting interference. Our following explanation is taken from [7] which is a thorough approach to the dirty paper coding problem that uses the THP framework. The concept can be easily explained for the case of one dimension as follows: Suppose we wish to transmit $u \in \left[-\frac{L}{2}, \frac{L}{2}\right)$ for some $L \in \mathbb{R}$ over the dirty paper channel given by (1). Let us forget about the additive noise $W$, for the moment, to establish the distinguishability of the signal $u$. Let the interference known at the transmitter be $s \in \mathbb{R}$. Instead of transmitting $u$, we intend to transmit $u - s$ so that the effect of the dirt is minimized. However, the power of $u - s$ might be very large and may not satisfy the transmit power constraint. The real trick in THP is to constrain the $u - s$ within the finite interval $\left[-\frac{L}{2}, \frac{L}{2}\right)$. Thus the transmitter transmits $u - s$ modulo-$L$. In other words, all the $u - s$ that differ by an integer multiple of $L$ are regarded as the same symbol. Now, in the absence of noise, the received signal is $u - s \mod L + s = u \mod L + (s - s \mod L) = u + (s - s \mod L)$. Clearly $[s - (s \mod L)] \mod L$ vanishes and even though the receiver does not know $s$, the receiver recovers $u$ by performing a modulo-$L$ operation. The utility of this approach is that the transmit signal is approximately uniformly distributed between $\left[-\frac{L}{2}, \frac{L}{2}\right)$, if the interference power is large, in which case the transmit power is $\frac{L^2}{12}$. Thus THP is a simple scheme that performs independent of the dirt power and does not compromise the distinguishability of the message. This approach is generalized to higher dimensions in [9] using a lattice vector quantizer and using a practical trellis precoding scheme in [7]. In [7], the designed scheme is shown to perform close to a trellis code on an AWGN channel for high signal to noise ratio (SNR) empirically. As we have been pointing out, in most applications of dirty paper coding, the integrity of dirt cannot be compromised. Let us focus on the behavior of dirt in the THP example. Observe that the receiver can only know $(s - s \mod L)$ and thus the THP method compromises the distinguishability of $s$ to achieve performance independent of dirt power. The main drawback of THP based methods is that it does *irrecoverable damage to the dirt* and this can prove costly when it is used in Gaussian broadcast channels and information embedding channels. We will, henceforth refer to such schemes as *dirt lossy*.

The channel described by (1) will be referred to, in general, as the dirty paper channel and the particular case of the channel described by (1) with zero dirt power will be called the corresponding interference free channel. While the corresponding interference free channel is just an AWGN channel, the term *corresponding* is used to denote that both the dirty paper channel and the corresponding interference free channel have the same Gaussian noise variance. In this paper, we consider the problem of finite constellation coding for the dirty paper channel with focus on reducing the probability of error for a given rate without affecting the dirt significantly. We mainly use Costa's work [1] to draw insights into designing codes for the dirty paper channel in this paper. Our comparison of the methods used for achievability of rates for the dirty paper channel with the methods used for achievability of rates for the AWGN channel motivates us to use a codeword, from the set of good partial interference presubtracted AWGN codewords, that is orthogonal to the given realization of the dirt. From the perspective of minimizing the pairwise probability of error, the goodness of an AWGN code is determined by the pairwise Euclidean distances between the codewords. Thus, if the pairwise Euclidean distances of a code are large, we may term the code a good AWGN code. It is well known that codes obtained by the trellis coded modulation (TCM) techniques introduced by Ungerboeck in [5] have large pairwise Euclidean distances. Thus, we design a scheme that chooses a TCM codeword (from a large bin of TCM codewords) that is almost perpendicular to the dirt. We find the dirt-orthogonal TCM codeword using the fast Viterbi algorithm. We empirically show that increasing the bin size (by increasing the number of states in the trellis and the spectral efficiency for which the trellis is designed) results in an increased coding gain.

The main contributions of this paper are:

- A design of a practical dirty paper coding scheme, for spectral efficiencies greater than 1 bit per channel use, whose construction mimics Costa's information theoretic proof and the performance of which is close to that of the AWGN channel.
- The observation that in most applications of dirty paper coding, like Gaussian broadcast channels, the dirt is actually valuable information and distorting the dirt to render good performance is not a good idea. Our designed scheme maintains the distinguishability of both the dirt and the transmit signal with high probability while the existing schemes (like lattice based methods [9],[10],[11] and THP based practical methods [7]) do not.

The rest of the manuscript is organized as follows: In Section II, we introduce the channel model. We explain Costa's information theoretic proof in Section III and derive finite constellation coding design insights. Section IV explains constructions of explicit finite constellation codes that satisfy derived criteria and illustrate the idea by an example. Section V addresses the applicability of our scheme to Gaussian broadcast channels. We discuss simulation results in Section VI. Finally Section VII concludes our paper by examining the deficiencies of this approach and noting the scope for future work.

*Notations:* Bold, uppercase letters are used to denote random vectors and bold, lowercase letters are used to denote vector realizations of the corresponding random variables. The set of all real numbers are denoted by $\mathbb{R}$. $\mathbb{E}(.)$ denotes expectation, $\Pr(.)$ denotes probability and $||.||$ denotes the Euclidean norm. For a real random variable $X$, $X \sim \mathcal{N}(0, \sigma^2)$ denotes that $X$ has a Gaussian distribution with mean 0 and variance $\sigma^2$. $Q(.)$ denotes the tail probability of the standard normal distribution, $Q(x) = \frac{1}{\sqrt{2\pi}} \int_x^\infty \exp\left(-\frac{u^2}{2}\right) du$.

## II. Model

Consider the problem of reliably communicating a message chosen uniformly from $\mathcal{I} = \{1, 2, 3, ..., Q\}$ over the discrete time channel shown in Fig. 1. Realizations of interference random variable $S \in \mathbb{C}$ are known non-causally at the transmitter and the dirty paper coding operation is performed to adapt the transmitted signal to the known channel interference. The random variables $W$ and $S$ are complex normal Gaussian with zero mean and variances $\sigma_w^2$ and $\sigma_s^2$ respectively. The random variables $W$, $S$ and $M$ are assumed to be independent. The message $M$ chooses $X(M, S) \in \mathbb{C}$ based on the known realization $S$ of the interference which is then transmitted over the channel. This constitutes using the dirty paper channel in one real dimension. We propose block transmission where transmission (and hence decoding) is in $n$ complex dimensions. The transmitted signal vector $\mathbf{x}(M, \mathbf{s}) \in \mathbb{C}^n$ is subject to an average power constraint (averaged over the messages), for a given vector realization $\mathbf{s} \in \mathbb{C}^n$, given by

$$\mathbb{E}(||\mathbf{X}(M, \mathbf{s})||^2) \leq P \qquad (2)$$

where, now, $M$ is the message chosen from the message set $\mathcal{I}$. The received signal vector $\mathbf{Y} \in \mathbb{C}^n$, according to the channel,

$$\mathbf{Y} = \mathbf{X}(M, \mathbf{S}) + \mathbf{S} + \mathbf{W} \qquad (3)$$

is used by the receiver to estimate $M$. The estimated message index is denoted by $\widehat{M}$.

## III. Costa's Proof and Insights to Finite Constellation Coding Design

### A. Costa's Proof

Costa computed the capacity of the channel shown in Fig.1 for the case of Gaussian dirt in [1]. With notations from section II, he showed that for any $R$ such that $R < C^* = \frac{1}{2} \log(1 + \frac{P}{\sigma_w^2})$ there exists length $n$ codes with $Q = 2^{nR}$ that achieve arbitrarily small probability of error for large $n$. And for all other $R$, such a code does not exist.

We urge the reader to refer to [1] to appreciate the following construction better. Observe that the above process involves constructing two codebooks $\tilde{\mathcal{C}}$ and $\mathcal{C}$. $\mathcal{C}$ is the actual codebook made up of points on the transmit sphere shown in fig. 2. The elements of the transmit sphere need to satisfy the transmit power constraint. An element of $\tilde{\mathcal{C}}$ is chosen and then modified according to the dirt after which it is transmitted (the modified vector, thus, belongs to $\mathcal{C}$). We will call this modified vector as a "transmitted codeword". Nevertheless we will call the elements of $\tilde{\mathcal{C}}$ also as codewords. It is important, for the reader, to bear in mind that a codeword and a transmitted codeword do not necessarily come from the same set. The code generation involves constructing a codebook $\tilde{\mathcal{C}}$ with large number of codewords whose co-ordinates are drawn from a zero mean Gaussian random variable with variance $P + \alpha^2 \sigma_s^2$. Each codeword, denoted by $\mathbf{U} \in \tilde{\mathcal{C}}$, is then assigned an index randomly from $\mathcal{I}$ and this process is termed *binning*. Note that this means we can have many codewords that have been assigned the same index and it may also happen that a certain index may not get assigned to any codeword. We will refer to a codeword that has been assigned an index $m \in \mathcal{I}$ as a "codeword from the bin $m$" and the set of codewords that have been assigned an index $m$ as "bin $m$" denoted by $B_m$. To transmit a message $m \in \mathcal{I}$, choose the bin $m$ and then select a codeword $\mathbf{U}$ from $B_m \subset \tilde{\mathcal{C}}$ that is jointly typical with $\mathbf{S}$. In essence, a codeword $\mathbf{U} \in B_m$ is chosen so that $|(\mathbf{U} - \alpha \mathbf{S})^T \mathbf{S}|$ is small enough. In other words, Costa's idea was to choose a transmitted codeword $\mathbf{X} = \mathbf{U} - \alpha \mathbf{S}$ that is, ideally, orthogonal to the dirt $\mathbf{s}$. A partial interference presubtraction (PIP) scheme refers to a scheme that transmits a signal that subtracts a fraction of the dirt vector from a codeword. PIP is performed in an attempt to reduce the effective Gaussian noise seen by the codeword. $\mathbf{X}$ is clearly a PIP codeword. The decoder just attempts to find a codeword $\mathbf{U}$ that is jointly typical with the received $\mathbf{y}$. It was proved in [1], that the rates achievable by this method are the same as that of the corresponding interference free channel by averaging over codebooks.

### B. Insights to Finite Constellation Coding Design

For the case of AWGN channel, the strategy of decoding using joint typicality of the transmit codewords and receive signal when transmit codewords were chosen from Gaussian generated codebooks achieved capacity of the AWGN channel. From a practical coding theoretic perspective, codebooks generated using TCM and decoding using minimum Euclidean distance decoder (MEDD) (which is actually a maximum likelihood decoder (MLD) for AWGN channel) yielded good performance.

Suppose we wish to communicate over the dirty paper channel at a spectral efficiency of $R$ and a transmit power $P$ using codewords of length $n$ and we know the dirt vector that will add is $\mathbf{s}$. As we know [1], the strategy to achieve capacity of the dirty paper channel involves two steps. The first step involves generating a Gaussian codebook $\tilde{\mathcal{C}}$ with a higher spectral efficiency (in this article, we will refer to it as *design spectral efficiency*) $\tilde{R} = R + R_0$ and a larger power (in this article, we will refer to it as *design power*) $\tilde{P} = P + \alpha^2 \sigma_s^2$. The codewords are then randomly binned into $2^{nR}$ partitions. The second step is to pick a bin corresponding to the message and search for a codeword sequence $\mathbf{u}*$ in the bin that is jointly typical with $\mathbf{s}$. The PIP codeword $\mathbf{x} = \mathbf{u}* - \alpha \mathbf{s}$ is then transmitted over the dirty paper channel. As explained by Costa, this operation corresponds to finding a $\mathbf{u}* \in \tilde{\mathcal{C}}$ so that $|(\mathbf{u}* - \alpha \mathbf{s})^T \mathbf{s}|$ is small. In other words, we want the transmit PIP codeword to be almost perpendicular the realization of the dirt vector. The decoder decodes a codeword from $\tilde{\mathcal{C}}$ that is jointly typical with the received signal. This strategy is proved to be rate optimal.

In what follows we design a practical dirty paper coding scheme by using the above idea. The MEDD turned out to be the probability of error minimizing decoder in the case of AWGN channel. Following Costa's idea and the idea that a codebook constructed by the TCM method has good distance properties, we will construct a TCM codebook $\tilde{\mathcal{C}}$ with elements drawn from a larger average power $\tilde{P}$ and the spectral efficiency of TCM codebook will be larger than $R$ to facilitate "binning". We will then partition these codewords at every node of TCM into $2^R$ bins (denoted by $B_m$ for $m \in \mathcal{I}$) and then choose a codeword $\mathbf{u} \in B_m$ so that the transmitted codeword $\mathbf{u} - \alpha \mathbf{s}$ (clearly a PIP codeword) has the smallest Euclidean inner product with the dirt $\mathbf{s}$ for some $\alpha \in \mathbb{R}$. A MEDD will be used at the receiver and the decoder will find the codeword from $\tilde{\mathcal{C}}$ that is the closest to the received vector $\mathbf{y}$. The receiver knows the partition of the code $\tilde{\mathcal{C}}$ and thus it finds the bin to which the decoded codeword belongs and decodes the bin index as the transmitted message.

The distinction between the transmitted codeword and codewords from $\tilde{\mathcal{C}}$ is now clear. The code $\tilde{\mathcal{C}}$ is the code used by the decoder to decode messages although what was transmitted was a vector that has modified the codeword from $\tilde{\mathcal{C}}$ to partially cancel the dirt and yet satisfy the transmit power constraint. It should be observed that for the receiver the channel looks like

$$\mathbf{y} = \mathbf{u} + (1 - \alpha)\mathbf{s} + \mathbf{w} \qquad (4)$$

which means while looking for $\mathbf{u}$ using a MEDD we are treating interfering fractional dirt as independent noise (Clearly this decoder is not a maximum likelihood decoder because the fractional dirt vector is not independent of the transmitted signal). This increases the effective noise seen by the codeword $\mathbf{u}$. Since the codeword $\mathbf{u}$ is drawn from a constellation with larger average power, the hope is that the receive signal to noise ratio (SNR) will be large enough yielding good performance.

Fig. 2 illustrates a scenario where we want to communicate at a spectral efficiency of 1 bit per channel use. The PIP codewords are within the sphere whose radius is $P$. The sphere is called the transmit sphere. The codewords on the transmit sphere are shaded differently based on the bin they belong. The dirt vector denoted by $\mathbf{s}$ is marked in the figure. Our aim is to find a codeword from a particular bin that is almost orthogonal to the given dirt.

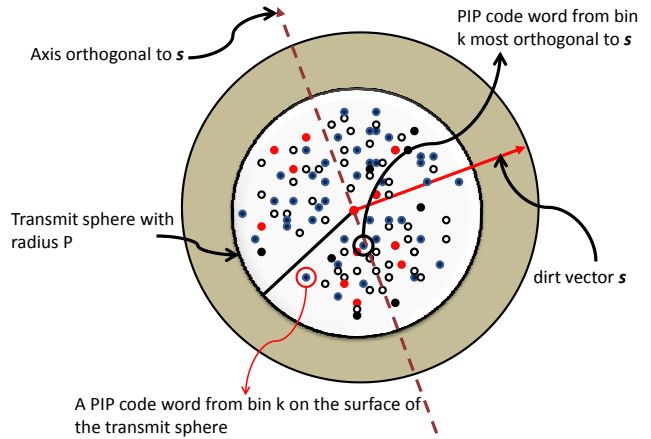

Fig. 2. An example outlining the method to choose PIP-TCM codewords when transmitting at unit spectral efficiency

## IV. DIRTY PAPER TRELLIS CODING

In this section we will describe the construction and decoding of a dirty paper coding scheme using finite constellation.

From the previous section we have inferred that we must design codes that satisfy the following design criteria:

1) Large pairwise Euclidean distance between the codewords for every realization of the dirt vector $\mathbf{s}$
2) Reduce the Euclidean inner product between the transmit PIP codeword and the known dirt vector $\mathbf{s}$

To satisfy the first criteria, we will use TCM to design our codes in the reminder of this section. Suppose we want to transmit messages at a spectral efficiency of $r \in \mathbb{R}^+$ bits per channel use and the average power constraint on the transmitter is $P$. The general idea is to construct a TCM based coding scheme for a design spectral efficiency $r_0 \in \mathbb{R}^+$ such that $r_0 > r$ and the signal points (denoted by $u$) are drawn from a finite constellation. A TCM scheme designed for a spectral efficiency of $r_0$ bits per channel use transmits one of $2^{r_0}$ different signal points for every channel use and this is represented by $2^{r_0}$ outgoing edges from each state in the trellis diagram. We partition the set of $2^{r_0}$ signal points from a state in the trellis into $2^r$ disjoint subsets (call it $\mathcal{B}_i$ for $i = 1, 2, 3 \cdots, 2^r$) so that each subset has $2^{r_0-r}$ signal points. The decoder is assumed to know the partitions. In our scheme, a message $m$ selects one of the $2^r$ subsets at each step and then the dirt vector $\mathbf{s}$ known to the transmitter helps choose a signal $u(m, \mathbf{s})$ among the $2^{r_0-r}$ possible choices. The knowledge of the entire non-causal realization of the dirt vector $\mathbf{s}$ is required to decide the codeword to be transmitted. In order to satisfy the second criteria with the method explained above for any realization of the dirt vector, we must decide a fast practical way to find a PIP codeword that is orthogonal to the dirt. The decoding is done by running a Viterbi algorithm and reconstructing a codeword. Then we identify the bin to which the codeword belongs by deciding the partition at every node.

The problem of finding a PIP-TCM codeword that is orthogonal to the given dirt is a hard problem to solve. We put down the problem mathematically:

**Problem:** Find a TCM codeword $\mathbf{u}$ from the chosen bin (obtained by a walk through the trellis) so that $|(\mathbf{u} - \alpha \mathbf{s})^H \mathbf{s}|$ is minimized.

**Basic Idea:** By triangle inequality,

$$|(\mathbf{u} - \alpha \mathbf{s})^H \mathbf{s}| < \sum_{i=1}^{i=n} |(u_i - \alpha s_i)^* s_i|$$

$$\min |(\mathbf{u} - \alpha \mathbf{s})^H \mathbf{s}| < \min \sum_{i=1}^{i=n} |(u_i - \alpha s_i)^* s_i|$$

Thus we try to search for a $\mathbf{u}^*$ that minimizes $\sum_{i=1}^{i=n} |(u_i - \alpha s_i)^* s_i|$ rather than the original Euclidean inner product. The problem of minimizing $\sum_{i=1}^{i=n} |(u_i - \alpha s_i)^* s_i|$ can be solved by running the Viterbi algorithm on the trellis. We use $|(u_i - \alpha s_i)^* s_i|$ as the branch metric and "decode" for a $u_i$ at each node among the TCM codewords corresponding to a particular bin.

Note that a codeword obtained by an arbitrary walk through the trellis is not necessarily a valid codeword in our setting. We require the message at each instant to choose one of the bins and only the signal points from the chosen bin at each node are possible contenders for presubtraction and then transmission. This motivates the following definition:

definition For a given message sequence $\mathbf{m} = (m_0, m_1, \cdots, m_n)$, a TCM codeword $\mathbf{u} = (u_0, u_1, \cdots, u_n)$ for which $u_i \in B_{m_i}$ is called a *valid TCM codeword*. definition In what follows, we illustrate our idea clearly by the means of an example.

Suppose that we want to transmit at a spectral efficiency of 1 bit per channel use, i.e. $r = 1$. Consider the 8 PAM constellation shown in fig. 4 and the corresponding trellis for the TCM scheme in fig 3. The numbers written on the left side of each node denote the labelling of edges so that they satisfy heuristic rules given in [5]. We will assume that the average power of the 8 PAM constellation, when all constellation points are equally likely, is the design power $\tilde{P}$. The TCM scheme, shown in fig. 3, generates codes with good Euclidean distance for the shown particular choice of set partitioning and labeling. We directly use the 8 state trellis, set partitioning and labeling used in the primary work [5]. The trellis shown in fig 3 constructs a TCM scheme for a design spectral efficiency of 2 bits per channel use i.e. $r_0 = 2$. We partition the edges into $2^r = 2$ sets of $2^{r_0 - r} = 2$ signal points each. In the figure, the signal points from the same set are depicted by either a dashed light line (corresponding to message 1) or a solid dark line (corresponding to message 0). When a message sequence and the dirt vector $\mathbf{s}$ are given, we use the Viterbi algorithm to find a path through the trellis whose PIP output is a valid codeword almost orthogonal to the dirt $\mathbf{s}$. While we are walking through the trellis, based on the message, we only consider the edges of a particular shade to be transmitted. Thus our $n$ length TCM code has $2^{2n}$ codewords which is partitioned into two sets at each node in the trellis giving $2^n$ bins having $2^n$ codewords each.

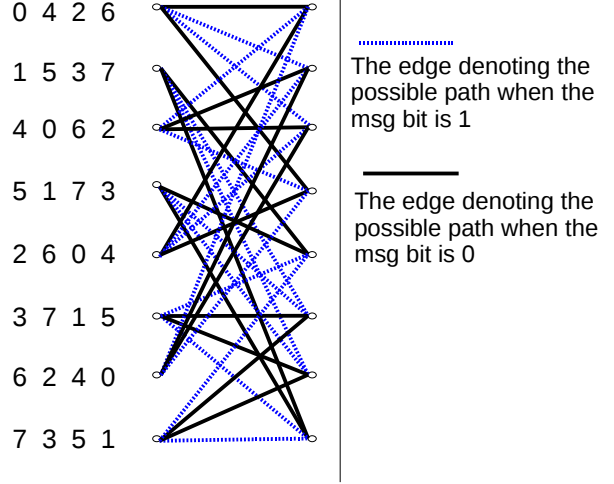

Fig. 3. TCM using 8 State Trellis and 8 PAM constellation for spectral efficiency of 2 bits per channel use

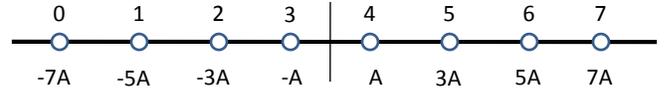

Fig. 4. Labelled 8 PAM constellation with $A = \sqrt{\frac{P}{21}}$

The idea behind the working of this scheme is the smart, effective use of PIP codeword. The PIP codeword helps in decreasing probability of error by reducing the net Gaussian noise power seen by the receiver. However the condition of orthogonality between the PIP codeword and the dirt helps in correlating the codeword $\mathbf{u}$ with the dirt so that the average power of the codebook containing $\mathbf{u}$ is increased (i.e. $\mathbb{E}(\|\mathbf{u}\|^2) = \tilde{P} > P$ ). Additionally, since we are traversing a subset of the edges at every node, the minimum Euclidean distance of our dirty paper trellis code will be higher than usual. Now we can see why our dirty paper code has the potential to match the performance of a TCM code on the corresponding interference free AWGN channel. An increased minimum Euclidean distance and an increased average receive power ($\tilde{P}$) try to compensate for the additional Gaussian noise, of power $(1 - \alpha)^2 \sigma_s^2$, seen by the signal $\mathbf{u}$.

## V. Applications to Gaussian Broadcast

The main aim of our design of dirty paper coding scheme is to make it applicable for Gaussian broadcast channels. As we have seen in the introduction, the dirt in the case of Gaussian broadcast channel is signal of the weaker user. We will now show why our scheme does not damage dirt in an information lossy way. Armed with the knowledge of the design explained in the previous section, we are in a position to appreciate why we can recover both the required signal and the dirt in our scheme.

*Theorem 1:* Our proposed scheme allows recovery of the dirt at the receiver with high probability.

*Proof:* We will to check whether the dirt $\mathbf{s}$ is recoverable. Observe that without noise the channel looks like

$$\mathbf{y} = \mathbf{u} + (1 - \alpha)\mathbf{s} \qquad (5)$$

Our receiver is designed to decode $\mathbf{u}$ with very low probability of error even when the additive noise is present. Clearly from (5), the channel, now, is an AWGN channel with lesser noise variance, from the perspective of the decoder. Since our receiver is employing MEDD and $\mathbf{u}$ is drawn from TCM codebooks, the receiver decodes $\mathbf{u}$ with an even lower probability of error than the case with noise. It should be clear that the probability of recovering $\mathbf{u}$ is the same as the probability of recovering $\mathbf{s}$. Our proposed scheme is designed to recover $\mathbf{u}$ with very high probability. Thus once the receiver recovers $\mathbf{u}$ (which happens with high probability), the receiver can subtract it from the received signal to recover $\mathbf{s}$. ∎

The TCM codebooks with Ungerboeck's method of set partitioning and labeling with sufficient expansion of the signal set gives high coding gain on the AWGN channel. In our case, if the dirt power is large then we will increase the overall spectral efficiency $r_0$ sufficiently high so that we will have more codewords in each bin. For a given transmit power constraint $P$, this will let us increase the average power of the constellation $\tilde{P}$. Thus the receiver will see codewords coming from a larger average power depending on the dirt power. Thus for any dirt power, the receive signal power (or the average power of the constellation from which the components of the codewords are drawn) increases accordingly, allowing us to recover both $\mathbf{u}$ and $\mathbf{s}$ with high probability.

We use our scheme to code for the broadcast channel, shown in fig. 5, in the following way: Suppose we want to transmit at a spectral efficiency of $b_s$ and $b_w$ bits per channel use to the strong user and weak user respectively. We first design a TCM codebook with a spectral efficiency of $b_w$ bits per channel use for the weak user. Then we use our dirty paper coding scheme to construct a PIP-TCM codebook, with a spectral efficiency of $b_s$ bits per channel use and sufficient signal set expansion. For the strong user so that the PIP-TCM codewords to be transmitted are as orthogonal as possible to the given codeword of the weak user. By using trellis shaping ([12] is a classic reference on trellis shaping) on the TCM codebooks of both the users, we can make the statistics of the component of the codewords look Gaussian. Thus each user's signal undergoes a noise that is effectively Gaussian, when a trellis shaping is used. Each user can then use a MEDD to decode their message reliably. The advantage of this method is that we need not reveal the code of one user to another. Such an arrangement can be useful when one needs to maintain secrecy of the message while broadcasting.

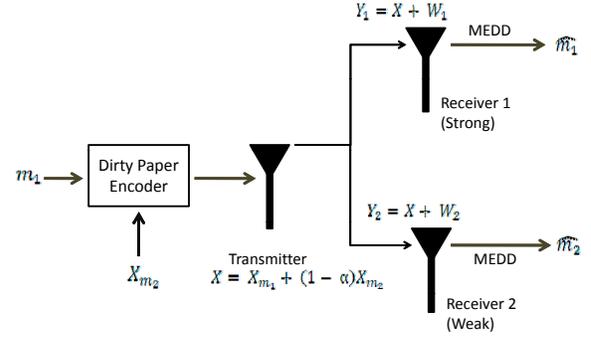

Fig. 5. A 2 user Gaussian broadcast channel employing dirty paper coding

A naive way to successfully communicate over a Gaussian broadcast channel is to choose zero mean constellations, $\mathcal{X}_1$ with average power $P_1$ and $\mathcal{X}_2$ with average power $P_2$, for each of user in such a way that their sum constellation $\mathcal{X}_1 + \mathcal{X}_2$ admits unique decodability. We say that the sum constellation has the property of unique decodability when each element of the sum constellation can be obtained by the addition of a unique combination of signal points, with one signal point chosen from each constellation (a necessary and sufficient condition for this is $|\mathcal{X}_1 + \mathcal{X}_2| = |\mathcal{X}_1|.|\mathcal{X}_2|$). This idea is actually a crude coding theoretic version of *superposition coding*. For the case of Gaussian MAC channel, this idea is necessary since the users transmit independently and there is no sum power constraint. One can use this type of "superposition coding" designed for Gaussian MAC for the case of broadcast (see [15], for instance). However, the performance is likely to be poor since we are not using the knowledge of the weak user's signal to gain any type of advantage (power or bandwidth). Clearly we can use joint decoding of both the symbols at both the receivers to decode our symbols. If $\mathbf{X_1} \in \mathcal{X}_1$ and $X_2 \in \mathcal{X}_2$ are chosen independently, then the average power of the sum constellation is

$$\mathbb{E}(|X_1 + X_2|^2) = \mathbb{E}(|X_1|^2) + \mathbb{E}(|X_2|^2) + 2\mathbb{E}(X_1^* X_2) \qquad (6)$$
$$= P_1 + P_2 \qquad (7)$$

In such an approach, the average power of the sum constellation is the sum of the average power of each constellation assuming the signal points are independently chosen. On the other hand, in our dirty paper coding approach, our aim in dirty

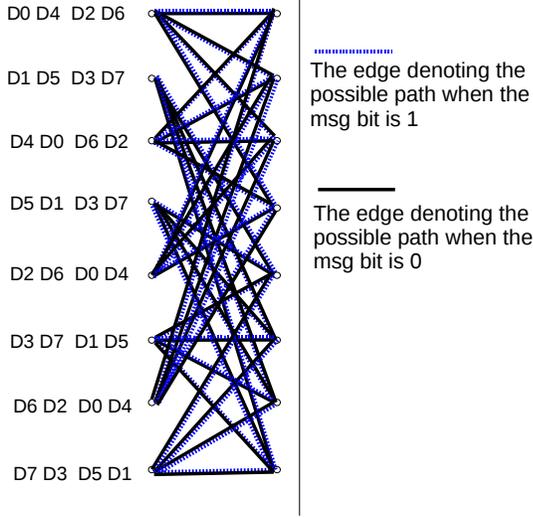

D0 D4  D2 D6

D1 D5  D3 D7

D4 D0  D6 D2

D5 D1  D3 D7

D2 D6  D0 D4

D3 D7  D1 D5

D6 D2  D0 D4

D7 D3  D5 D1

········ The edge denoting the possible path when the msg bit is 1

——— The edge denoting the possible path when the msg bit is 0

Fig. 6.   Dirty paper TCM using 8 State Trellis for 16 QAM constellation indicating the used set partition

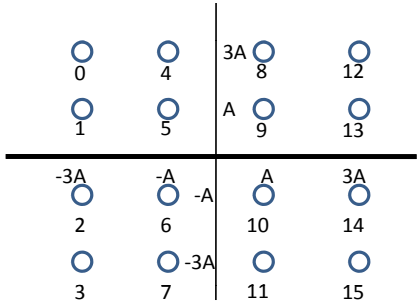

Fig. 7.   Labelled 16 QAM constellation $A = \sqrt{\frac{P}{10}}$

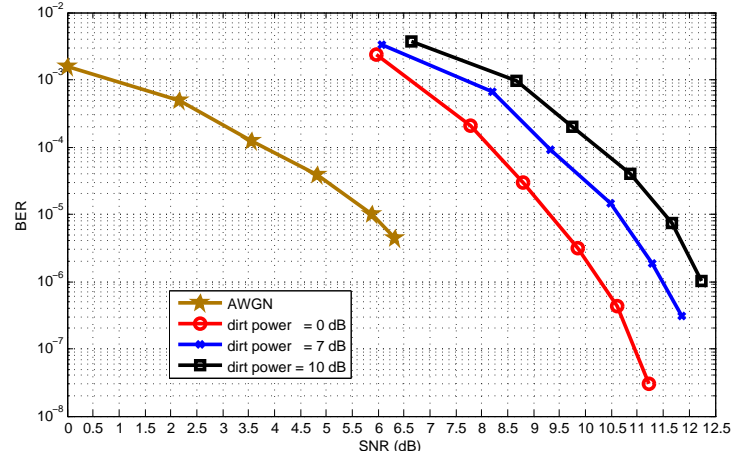

Fig. 8.   BER versus SNR at 1 bit/channel use using a 8 state trellis and 16 QAM constellation with alpha=0.9

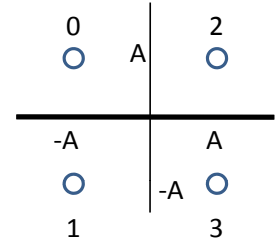

Fig. 9.   Labelled 4 QAM constellation $A = \sqrt{\frac{P}{2}}$

paper coding is to reduce $|(U - \alpha S)^H S|$ as much as possible. In our design of dirty paper coding, we have explained the idea of increasing spectral efficiency to increase the size of the bin to increase $\tilde{P}$. From Costa's proof, we know that $\tilde{P} = P + \alpha^2 \sigma_s^2$ for the limiting case of infinite alphabet with Gaussian distribution. For the sake of analysis, let us assume $(\mathbf{u} - \alpha \mathbf{s})^H \mathbf{s} \approx 0$. Thus we have

$$\mathbf{u}^H \mathbf{s} \approx \alpha ||\mathbf{s}||^2 \tag{8}$$

The codewords in our scheme are aligned to make them statistically correlated. This allows the strong user to have an increased design power $\tilde{P}$ for the transmit average power of $P$. Thus using dirty paper coding boosts the strong user's effective power using the correlation with the weak user's signal. Superposition coding loses to dirty paper coding in harvesting this power advantage.

## VI. SIMULATION RESULTS

The bit error rate (BER) was obtained by simulating our scheme in blocks of length $N = 10^5$ until at least 100 errors accrued. The spectral efficiency was fixed at 1 bit per channel use. Thus for each iteration, a block of equally likely bits of length $N$ were generated as the message sequence. The dirt and the message sequence selected a PIP-TCM codeword using a Viterbi algorithm at the encoder. SNR was measured as ratio of power of the simulated complex transmit signal to power of the simulated complex noise power. We remark that we have not used trellis shaping [12] which could have resulted in an additional gain.

Fig. 8 depicts the simulation performance for case of an 8 state trellis, shown in fig. 6 (the numbers written on the left side of each node denote the set partition so that they satisfy heuristic rules given in [5]) with 16 QAM constellation, shown in fig. 7. The set partitioning and labelling followed are directly used from [5] where D0 = $\{0, 10\}$, D1 = $\{6, 12\}$, D2 = $\{5, 15\}$, D3 = $\{3, 9\}$, D4 = $\{8, 2\}$, D5 = $\{4, 14\}$, D6 = $\{7, 13\}$, D7 = $\{1, 11\}$. The noise power was fixed as $\sigma_w^2 = 1$ and the design power $\tilde{P}$ was varied. While performing PIP, $\alpha$ was chosen to be 0.9 in this case. For a BER = $10^{-5}$, with $\sigma_s^2 = 10$ the SNR required is 11.6dB, with $\sigma_s^2 = 5$ the SNR required is 10.6dB, with $\sigma_s^2 = 1$ the SNR required is 9.2dB. This shows that the BER performance improves when the dirt power decreases when the value of $\alpha$ is fixed. The

## 4 State Trellis

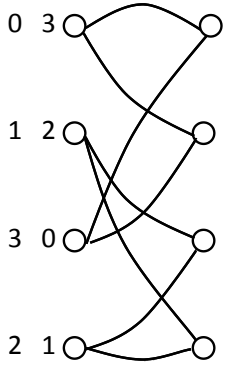

Fig. 10.  TCM using 4 State Trellis for 4 QAM constellation indicating the used set partition

## 4 State Trellis

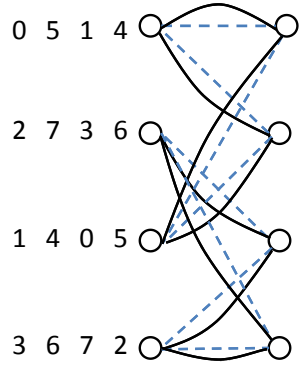

The edge denoting the possible path when the message bit is 0

The edge denoting the possible path when the message bit is 1

Fig. 12.  Dirty paper TCM using 4 State Trellis for 8 QAM constellation indicating the used set partition

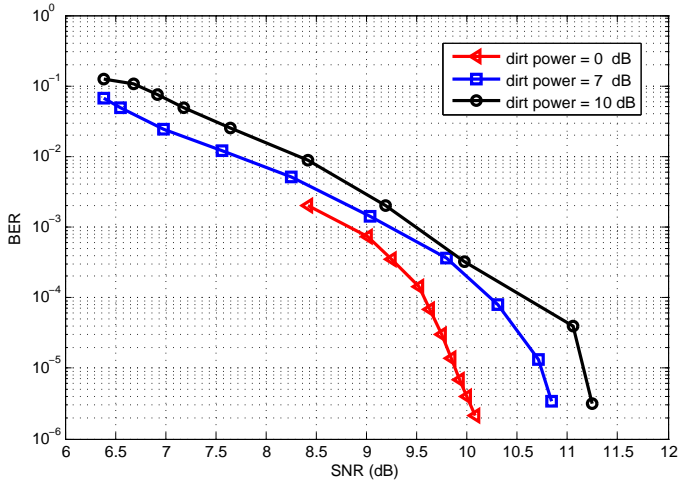

Fig. 11.   BER versus SNR at 1 bit/channel use using a 8 state and 16 QAM constellation with $\tilde{P} = 16.3$ dB

## 8 TRELLIS STATES

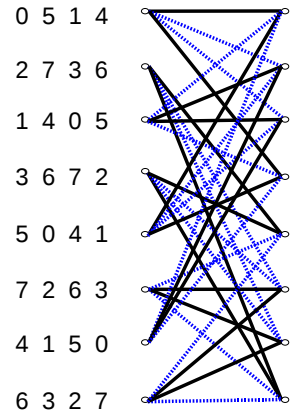

The edge denoting the possible path when the msg bit is 1

The edge denoting the possible path when the msg bit is 0

Fig. 13.   Dirty paper TCM using 8 State Trellis for 8 QAM constellation indicating the used set partition

AWGN performance shown in fig. 8 has a BER = $10^{-5}$ at a SNR of 5.9dB and thus our scheme with $\sigma_s^2 = 1$ is within 3 dB of AWGN performance. The equivalent AWGN code is a TCM code designed for a spectral efficiency of 1 bit per channel use using a 4 QAM constellation, shown in fig. 9, using a 4 state trellis shown in fig. 10. We remark that our scheme performs well since even the "dirt lossy" THP based trellis precoding scheme constructed in [7] is around 2.5dB away from AWGN performance with an equivalent code for a spectral efficiency of 1 bit per channel use.

Fig. 11 depicts the simulation performance for case of an 8 state trellis, shown in fig. 6 with 16 QAM constellation, shown in fig. 7 with a fixed design power $\tilde{P} = 16.3$dB. The noise power was fixed as $\sigma_w^2 = 1$ and $\alpha$ was varied from 0 to 1 in steps of 0.1. This plot explores the effect of varying $\alpha$. For a fixed design power $\tilde{P}$, as $\alpha$ is varied, the transmit power varies. The change in transmit power is reflected in the measured SNR. Clearly both the design power and $\alpha$ are design parameters. As clearly seen from fig. 11, for a particular BER (say $10^{-5}$), with $\sigma_s^2 = 10$ the SNR required is 11.2dB, with $\sigma_s^2 = 5$ the SNR required is 10.7dB, with $\sigma_s^2 = 1$ the SNR required is 9.7dB. Thus the BER performance improves when the dirt power decreases when the value of design power $\tilde{P}$ is fixed. Also, the relation between design power $\tilde{P}$ and transmit power $P$ is unknown analytically and depends on the number of states and the design spectral efficiency $r_0$.

The overall performance improves as we increase the number of states and increase the design spectral efficiency $r_0$. This effect is clearly seen in fig. 15. We observe, from fig. 15, that a 4 state trellis, shown in fig. 12 with 8 QAM constellation, shown in fig. 14 achieves a BER = $10^{-5}$ with $\sigma_s^2 = 1$ for a SNR of 10.8dB. An 8 state trellis, shown in fig. 13 with 8 QAM constellation, shown in fig. 14 achieves a BER = $10^{-5}$ with $\sigma_s^2 = 1$ for a SNR of 10.2dB. An 8 state trellis, shown in fig. 6 with 16 QAM constellation, shown in fig. 7 achieves a BER = $10^{-5}$ with $\sigma_s^2 = 1$ for a SNR of 9.5dB. Thus it is clear that increasing the number of states and increasing the design spectral efficiency improves performance by increasing the obtained coding gain.

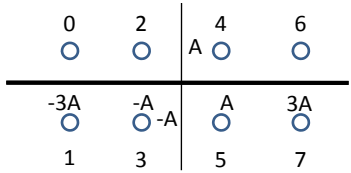

Fig. 14. Labelled 8 QAM constellation $A = \sqrt{\frac{P}{6}}$

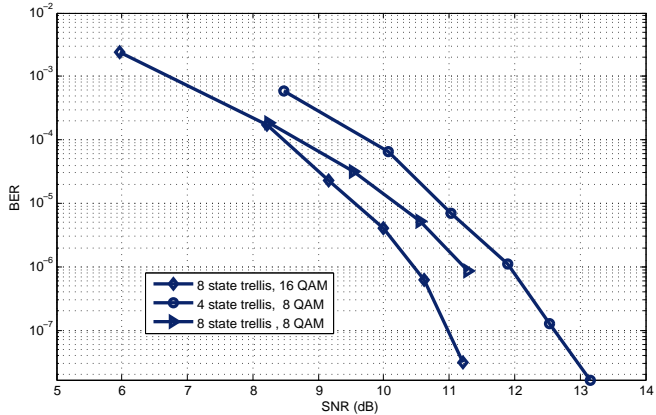

Fig. 15. The coding gain obtained as we increase the spectral efficiency of the TCM scheme and the number of states

## VII. DISCUSSION

We have presented a novel dirty paper coding scheme which followed Costa's paper analytically and allowed the recoverability of the dirt. Our scheme implemented random binning in a systematic manner using a trellis and used Viterbi algorithm to quickly obtain a PIP-TCM codeword that is almost orthogonal to the realization of the dirt. We empirically demonstrated that this scheme performs close to an equivalent TCM code on AWGN channel. We also showed that coding gain increases as design spectral efficiency and number of states are increased. A few shortcomings of our scheme that can prove useful for future research is the following:

- We need the entire message sequence to construct the PIP-codeword for transmission. While using large codeword lengths is profitable in terms of error performance, the communication incurs a huge delay while encoding. A way to remedy this is to truncate the survivors to some manageable length in the Viterbi algorithm as indicated in [16]. If the survivor paths at some time had a common tail until a particular depth, then we can output the "decoded" symbols. However if the survivors disagree on the initial path then a good strategy for determining the initial path is not very clear.

- Our scheme has implemented random binning and searched for a codeword in a bin using the Viterbi algorithm. We have not designed a criteria for partitioning the signal sets. Since it was seen that certain partitions behave differently than the others, it might be instructive to investigate the problem of partitioning signal points at the nodes to enhance error performance.

- For Gaussian codebooks, the relation between the design power $\tilde{P}$ and transmit power $P$ is concrete, i.e. $\tilde{P} = P + \alpha^2 \sigma_s^2$ and so is the optimal $\alpha = \frac{P}{P + \sigma_w^2}$. However for finite constellation, all we can say is $\tilde{P} > P$ and the optimal $\alpha$ depends on the dirt power. The relation between design power $\tilde{P}$ and transmit power $P$ is not clear analytically. We only know that it depends on the chosen constellation and the chosen trellis. It will be very useful in designing dirty paper TCM codes if the relation between them was characterised analytically. An exact characterisation or even a relation in terms of bounds on the design power will prove useful.

Information theoretic designs are impractical because decoding techniques like joint typical decoding require a brute force search. Our work interpreted the meaning of typicality to obtain a nice criteria and then we managed to construct a fast algorithm to satisfy the criteria, thereby designing a good scheme for the dirty paper coding problem. The message of this paper is that, while information theoretic approaches that search for codewords in a random bin might seem like a hard problem, slight engineering approximations might lead to an easier problem for which a fast algorithm exists. This paper introduces *a novel approach* to coding theory design by additionally using the criteria from information theoretic viewpoints rather than focus on just traditional coding theoretic concepts like deriving design criteria by minimizing pairwise error probability.

## ACKNOWLEDGEMENT

This work was partly supported by the DRDO-IISc program on Advanced Research in Mathematical Engineering, through a research grant to B.S. Rajan.

## REFERENCES

[1] M. Costa, "Writing on dirty paper (corresp.)", *IEEE Trans. Inf. Theory*, vol. 29, no. 3, pp. 439-441, 1983.

[2] M. Tomlinson, "New automatic equalizer employing modulo arithmetic", *Electronic Letts.*, vol. 7, pp. 138-139, Mar. 1971.

[3] H. Miyakawa and H. Harashima, "Information transmission rate in matched transmission systems with peak transmitting power limitation", *Nat. Conf. Rec. Inst. Electron. Inform. Commun. Eng. Japan*, vol. 7, no. 2, pp. 138-139, Aug. 1972.

[4] S. I. Gel'fand and M. S. Pinsker, "Coding for channel with random parameters", *Prob. Cont. Inform. Theory*, vol. 9, no. 1, pp. 19-31, 1980.

[5] G. Ungerboeck, "Channel coding with multilevel/phase signals", *IEEE Trans. Inf. Theory*, vol. IT-28, no. 1, pp. 55-67, Jan. 1982.

[6] W. Yu and J. M. Cioffi, "Trellis precoding for the broadcast channel", *Proceedings of 2001 IEEE Global Telecommunications Conference*, vol. 2, pp. 1344-1348, Nov. 2001.

[7] W. Yu, D. P. Varodayan, J. M. Cioffi, "Trellis and convolutional precoding for transmitter based interference presubtraction", *IEEE Trans. Commun*, vol. 53, no. 7, pp. 1220-1230, 2005.

[8] B. Chen and G. W. Wornell, "Quantization Index Modulation: A Class of Provably Good Methods for Digital Watermarking and Information Embedding", *IEEE Trans. Inf. Theory*, vol. 47, no. 4, pp. 1423-1443, 2001.

[9] U. Erez, S. Shamai, and R. Zamir, "Capacity and Lattice strategies for cancelling known interference", *Proc.Int.Symp.Inf.Theory Applicat.*, pp. 681-684, Nov. 2000.

[10] R. Zamir, S. Shamai, and U. Erez, "Nested linear/lattice codes for structured multiterminal binning", *IEEE Trans. Inf. Theory*, vol. 48, no. 6, pp. 1250-1276, 2002.

[11] U. Erez and S. Brink, "A close-to-capacity dirty paper coding scheme", *IEEE Trans. Inf. Theory*, vol. 51, no. 10, pp. 3417-3432, 2005.

[12] G. D. Forney, "Trellis Shaping", *IEEE Trans. Inf. Theory*, vol. 38, no. 2, pp. 281-298, Mar. 1992.

[13] M. Skoglund and E.G. Larsson, "Optimal modulation for known interference", *IEEE Trans. Inf. Theory*, vol. 56, no. 11, pp. 1892-1899, 2008.

[14] T. Gariby, U. Erez, S. Shamai, "Dirty Paper Coding with a Finite Input Alphabet", *Proceedings of 2006 IEEE 24th Convention of Electrical and Electronics Engineers in Israel*, pp. 95-99, Nov. 2006.

[15] J. Harshan and B. S. Rajan, "Coding for Two-User Gaussian MAC with PSK and PAM Signal Sets", *Proceedings of 2009 Int.Sym.Inf.Theory*, Seoul, Jun. 2009.

[16] G. D. Forney, "The Viterbi Algorithm", *Proc. IEEE*, vol. 61, no. 3, pp. 268-278, Mar. 1973.